\documentclass[prl,twocolumn, floatfix]{revtex4-1}
\usepackage{graphicx}
\usepackage{epsfig}
\usepackage{verbatim}
\usepackage{color}

\begin{document}

\title{Scalable ultra-sensitive detection of heterogeneity via coupled bistable dynamics}

\author{Kamal P. Singh\thanks{e-mail: kpsingh@iisermohali.ac.in}, Rajeev
Kapri\thanks{e-mail:rkapri@iisermohali.ac.in} and Sudeshna Sinha\thanks{e-mail:
sudeshna@imsc.res.in (On leave from Institute of Mathematical Sciences, Chennai 600113,
India)}}

\affiliation{Indian Institute of Science Education and Research (IISER) Mohali, SAS Nagar,
Sector 81, Mohali 140 306, Punjab, India}

\begin{abstract}

	We demonstrate how the collective response of $N$ globally coupled bistable elements can
	strongly reflect the presence of very few non-identical elements in a large array of
	otherwise identical elements. Counter-intuitively, when there are a small number of elements
	with natural stable state different from the bulk of the elements, {\em all} the elements of
	the system evolve to the stable state of the minority due to strong coupling. The critical
	fraction of distinct elements needed to produce this swing shows a sharp transition with
	increasing $N$, scaling as $1/\sqrt{N}$. Furthermore, one can find a global bias that allows
	robust {\em one bit} sensitivity to heterogeneity. Importantly, the time needed to reach the
	attracting state does not increase with the system size.  We indicate the relevance of this
	ultra-sensitive generic phenomenon for massively parallelized search applications.

\end{abstract}

\maketitle

The complex interactive systems modelling spatially extended physical, chemical and biological
phenomena, has commanded intense research effort in recent years. One of the important issues
in such systems is the effect of heterogeniety on spatiotemporal patterns. The role of
disorder, such as static or quenched inhomogenieties, and the effect of coherent driving
forces, has yielded a host of interesting, often counter-intuitive, behaviours. For instance,
stochastic resonance \cite{sr} in coupled arrays \cite{spatial_sr}, diversity induced resonant
collective behaviour in ensembles of coupled bistable or excitable systems \cite{diversity}
demonstrated how the response to a sub-threshold input signal is optimized. However, the
potential of coupled system for parallel information processing remains largely unexplored. It
is our aim to show the existence of an ultra-sensitive regime of $N$-coupled bistable elements
whereby small heterogeneity in the system strongly influence its global dynamics.

Here we consider $N$ coupled nonlinear systems, where the evolution of element $i$ ($i =1,
\dots N$) is given by:
\begin{equation}
  \dot{x_i} = F(x_i) + a_i + C (\langle x \rangle -x_i) + b
\label{array}
\end{equation}
where $C$ is the coupling strength, $b$ is a small global bias, and the mean field $\langle x
\rangle$ is given by
\begin{equation}
\langle x \rangle = \frac{1}{N} \sum_{i=1,N} x_i
\label{xav}
\end{equation}
The function $F(x)$ is nonlinear and yields a bistable potential.  Specifically, we consider
$F(x) = x_i -x_i^3$, which gives rise to a double-well potential, with one well centered at
$x_-^* = -1$ (lower well) and another at $x_+^* = +1$ (upper well).

We consider a situation where $a_i$ can take either of two sufficiently different values, $A_0$
and $A_1$. With no loss of generality, we set $A_0 =0$ and $A_1 = 1$, i.e.  $a_i$ of the
elements $i = 1, \dots N$, can be $0$ or $1$. The initial conditions on $N$ elements here are
taken to be randomly distributed about zero mean, and in our simulations a very large number of
initial states ($\sim 10^4$) are sampled in different trial runs. In order to quantify what
fraction of elements have $a_i=0$, we use the following notation: $N_0$ is the number of
elements with $a_i = 0$, and $N_1 = N - N_0$ is the number of elements with $a_i = 1$. The
principal question is: how sensitive are collective dynamical features, such as the ensemble
average $\langle x \rangle$, which can be considered as the {\em output} of the system, to
small inhomogeneity [see Fig. 1(a)]. 

The values $A_0$, $A_1$ and bias $b$ are such that in an uncoupled system, when $a_i = A_0$,
the system goes rapidly to the lower well $x^*_{-} \sim -1$, while the system with $a_i = A_1$
is attracted to the upper well $x^*_{+} \sim 1$. When all $a_i = A_0$, we have a homogeneous
system, and this uniform system is naturally attracted to the lower well $x^*_{-}$. One may
wonder, how many $a_i$ need to be different from $0$ in order to make a significant difference
in the collective output.

Intuitively, one may expect that the global average will pick up contributions of order $1/N$
from each element. So a fairly large number of elements need to be different in order to obtain
significant deviation in the mean field and drive a different collective response. Alternately,
for strong coupling, one may think, for small heterogeneity, the majority of the elements will
dictate the nature of the collective field, as the minority should synchronize with the
majority population.

However, we will show here that both the expectations above do not hold true. Instead, this
system, under sufficiently strong coupling, will evolve to the stable state of the {\em
minority population}. Furthermore, the critical number of elements distinct from the bulk that
is needed for this effect, is typically less than $O(\sqrt{N}$), and can actually be made as
small as {\em one}.

Figure 1(b) shows the evolution of $100$ globally coupled elements, with coupling constant $C=1$,
to the {\em lower} well from Gaussian random initial conditions, in the homogeneous case where
all $a_i =0$. In sharp contrast, Fig. 1(c) shows all the elements of the array being attracted
to the {\em upper} well, when a few elements have $a_i =1$. So it is clearly evident that even
when very few $a_i$s are different from $0$, the entire array is pushed to the upper well.

\begin{figure}[t]
\begin{center}
\includegraphics[width=0.9\linewidth]{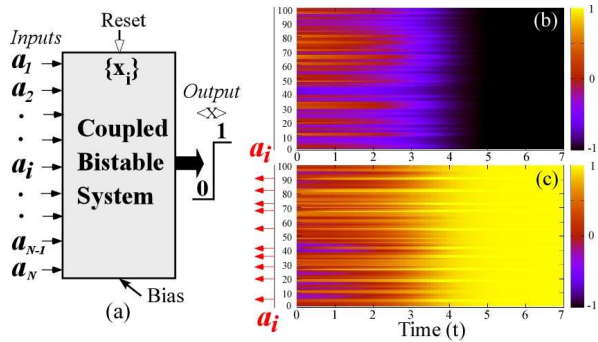}
\end{center}
\caption{(a) Schematics of coupled nonlinear system with N-inputs, output, bias, and reset for
initial conditions. (b) time evolution of an array of $100$ strongly coupled ($C=1$) bistable
elements having {\em all} $a_i=0$ in Eq.(1), and (c) having $10$ elements with $a_i=1$ (denoted
by arrows) and $90$ elements with $a_i=0$.}
\label{schema}
\end{figure}

So the collective field sensitively reflects very small deviations from uniformity. In fact,
the response to small diversity is a swing from the lower well to the upper one, for {\em all}
elements in the system. The coupled system then acts like a sensitive detector, as its response
to few $a_i = 1$, in an otherwise uniform lattice of $a_i=0$, is very large.

Coupling is crucial in this effect. In a weakly coupled system, when a few $a_i$ are different,
the difference in the mean field of the homogeneous and inhomogeneous systems will be
proportional to $N_1/N$. However, the response of the strongly coupled system to small $N_1$ is
very large (namely $\sim (x^*_{+} - x^*_{-})$).

In order to quantify the sensitivity of the collective response, we calculate the minimum $N_1$
needed to flip the output to the upper well (within a small prescribed accuracy). We call this
the {\em critical population $N_{1c}$}.

\begin{figure}[t]
  \begin{center}
    \includegraphics[width=0.7\linewidth]{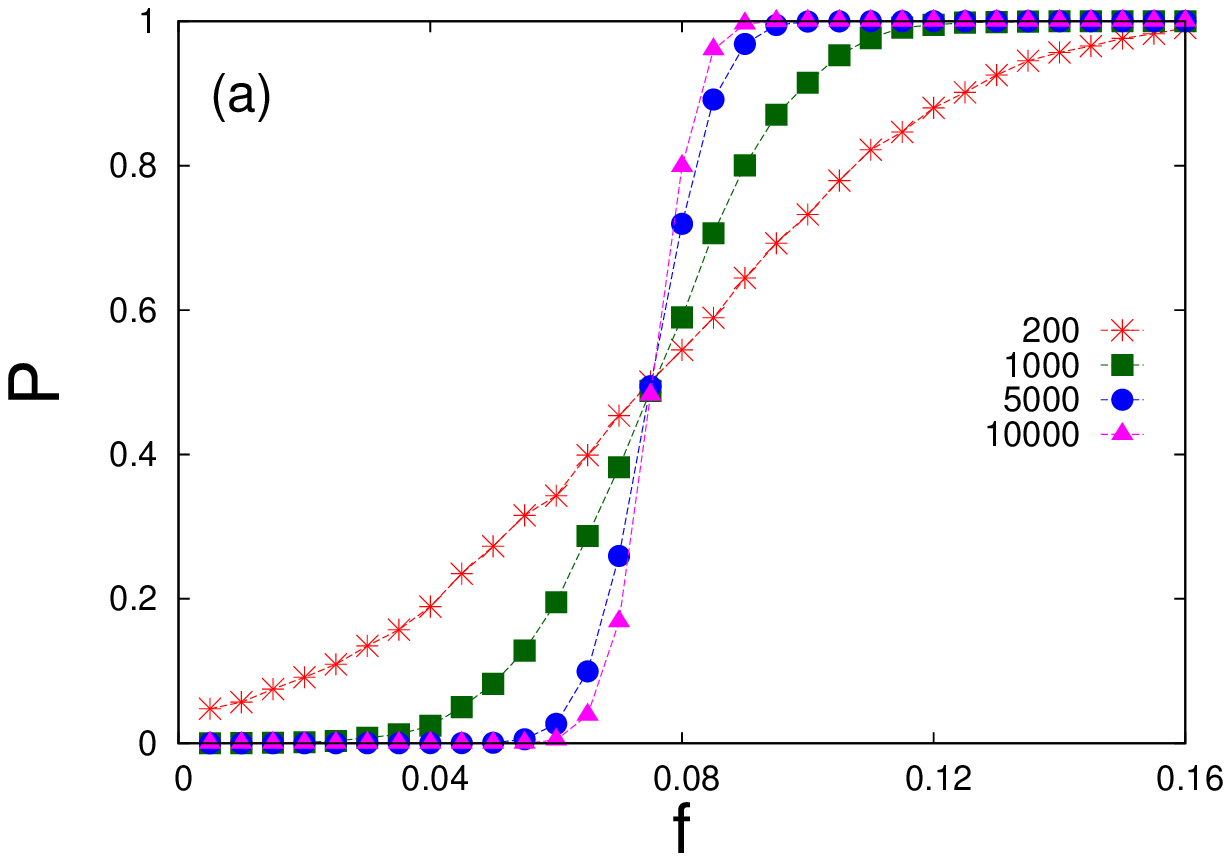}
    \includegraphics[width=0.7\linewidth]{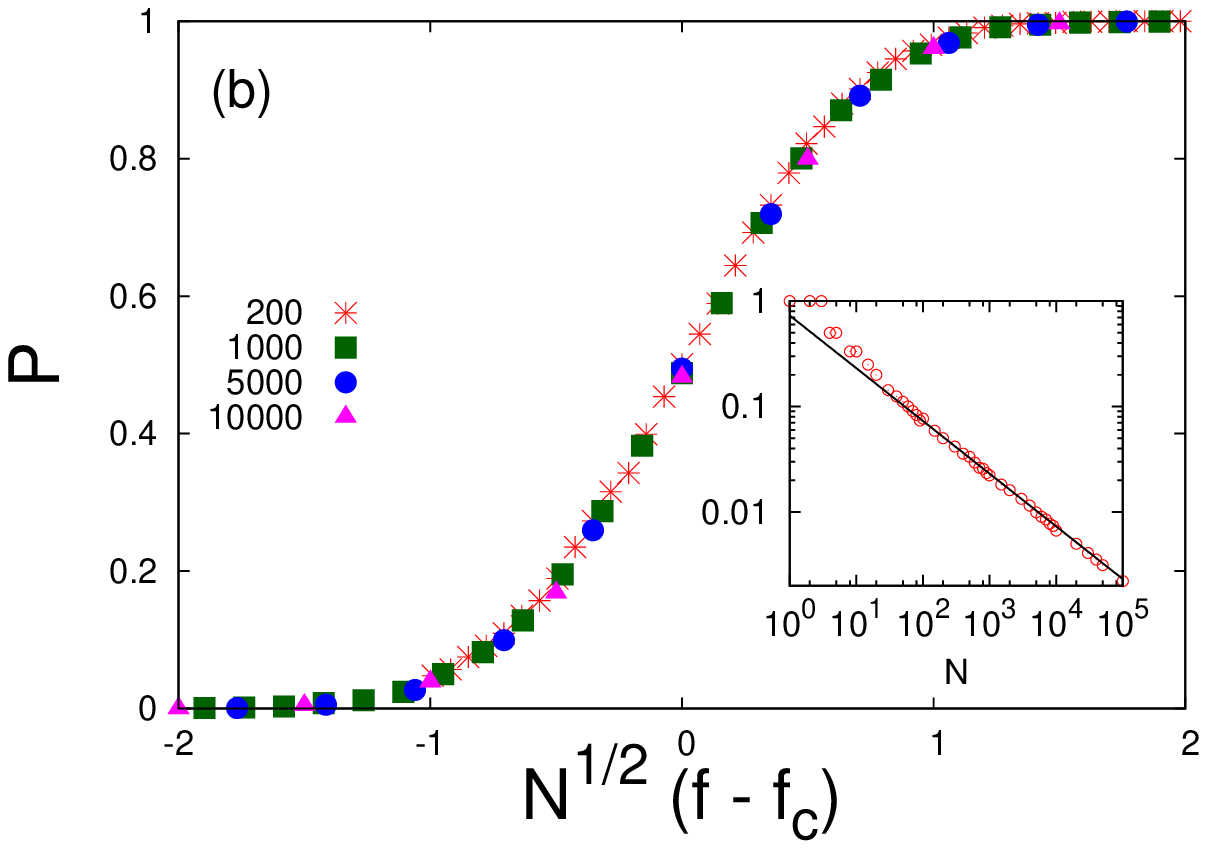}
		\includegraphics[width=0.7\linewidth]{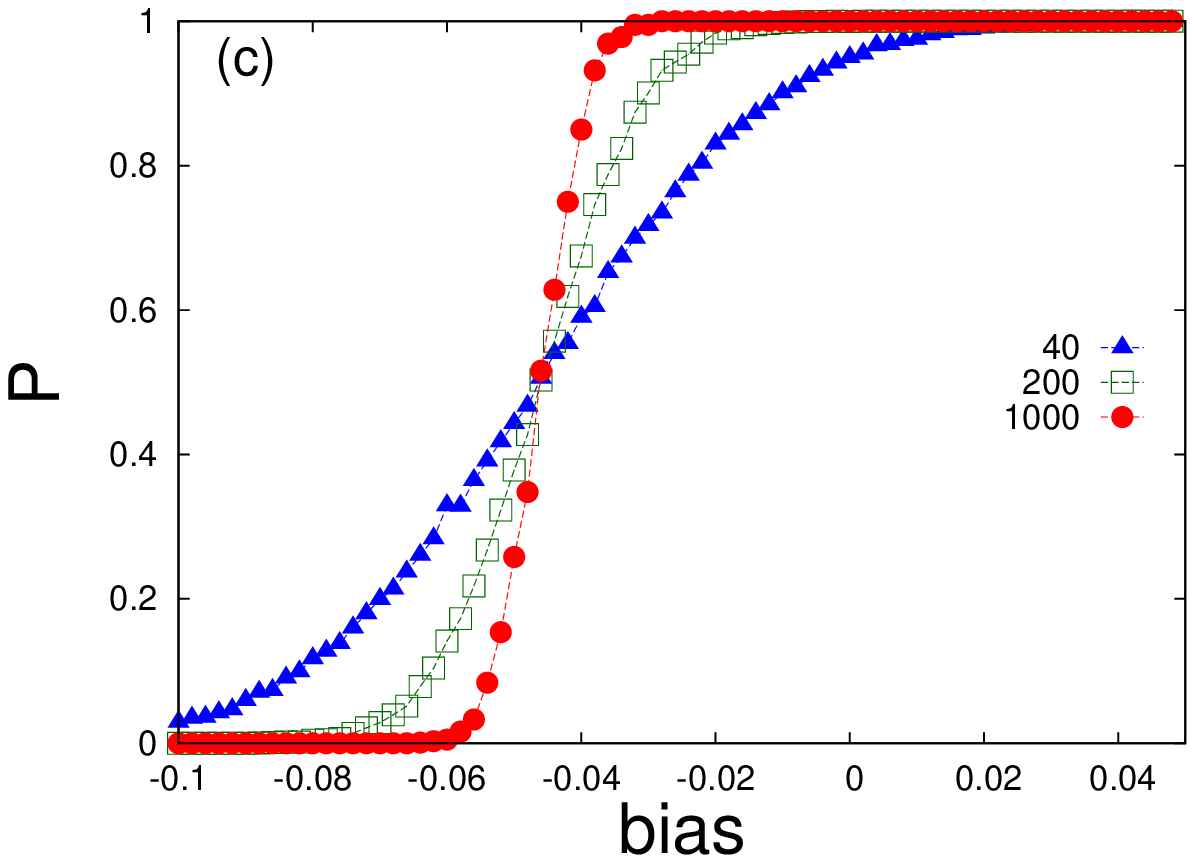}
  \end{center}
	\caption{(a) Probability, $P$, that all the elements evolve to the upper well, from random
	initial conditions, as a function of fraction $f=N_1/N$ for different system sizes. Here bias
	$b=-0.05$, coupling constant $C=1$, and averaged over $10,000$ runs. (b) Data collapse of
	probability $P$ for various $N$ as obtained by scaling the $x$-axis by $N^{1/2}(f-f_c)$,
	indicates that $f_c$ is $0.075 \pm 0.001$ for this bias. Inset: scaling of $f-f_c$ with
	respect to $N$, where $f$ is the fraction at which $P \sim 1$; solid line: $1/\sqrt{N}$ fit.
	(c) $P$ vs global bias $b$, for different system sizes, with $f=N_1/N=0.075$.}
 	\label{scaling}
\end{figure}

Figure \ref{scaling} displays the collective response to very small inhomogeneity under varying
system size $N$ and global bias $b$. In Fig.~\ref{scaling}(a), we show the probability, $P$,
that all the oscillators evolve to the upper well, as a function of $f = N_1/N$ for different
system sizes for a bias $b=-0.05$ and coupling constant $C=1$.  The figure shows that there is
a critical fraction above which all the oscillators switch to the upper well. The switching
becomes sharper and sharper as the system size increases. To obtain the critical population,
$N_{1c}$ in the large $N$ limit, we use the finite size scaling. For a given bias and coupling
constant, the probability $P$ satisfies the following scaling form
\begin{equation}
  P \sim {\mathcal G} \left( N^{\phi} (f - f_{c}) \right),
  \label{eq:fss}
\end{equation}
where $f_c = N_{1c}/N$ when $N \rightarrow \infty$, $\phi$ is the critical exponent and
${\mathcal G}$ is the scaling function. A good data collapse, shown in Fig.~\ref{scaling}(b),
is obtained for $\phi = 1/2$ indicating that
\begin{equation}
  \left| f - f_{c} \right| \sim \frac{1}{\sqrt{N}} \Rightarrow  \left| N_1
  - N_{1c} \right| ~\sim \sqrt{N}.
  \label{critical}
\end{equation}
Clearly then, the minority population can pull the strongly coupled bistable system to a final
state distinct from the homogeneous case.

\begin{figure}[t]
  \begin{center}
    \includegraphics[width=0.475\linewidth,angle=270]{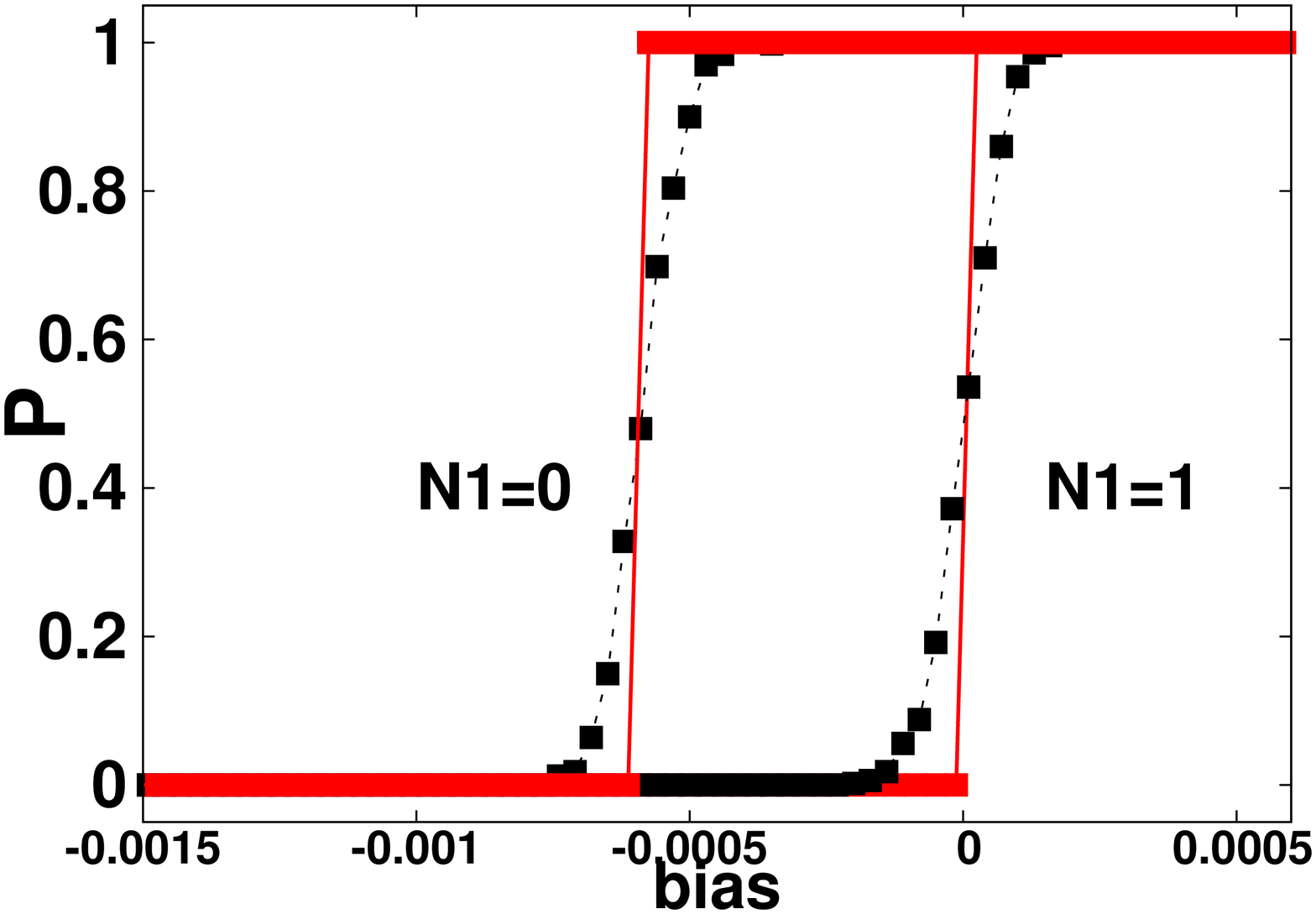}
    \includegraphics[width=0.75\linewidth]{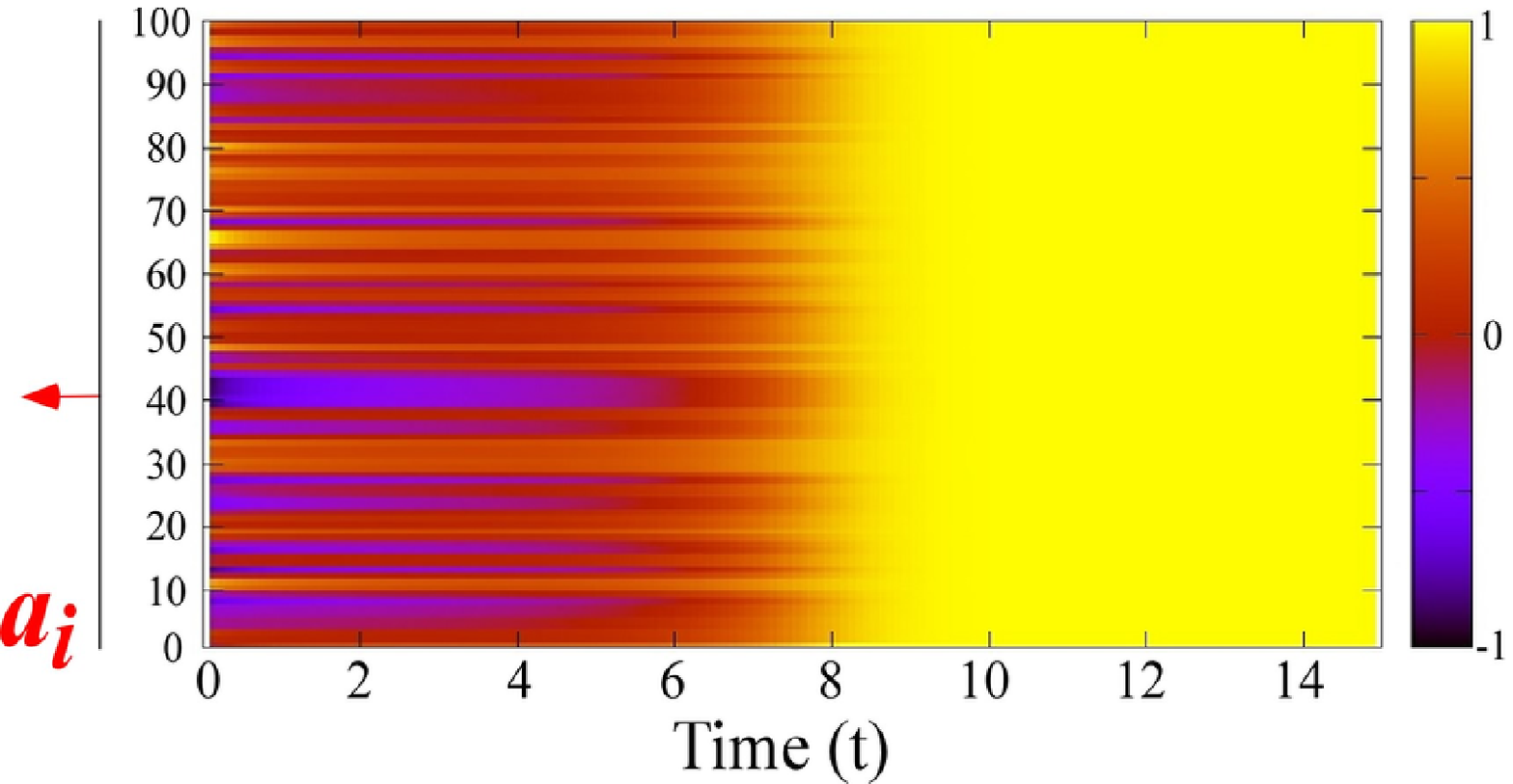}
  \end{center}

	\caption{ (Top Panel) Probability, $P$, that all the oscillators evolve to the upper well as
	a function of global bias, for a system of $1000$ elements: (right curve) with no $a_i=1$,
	namely $f=N_1/N=0$ and (left curve) with just {\em one} element with $a_i=1$, namely $f=
	N_1/N = 0.001$ (left curve). The red (solid) line shows the deterministic case with all
	initial $x = 0$. The green (dashed) curve shows the evolution from an initial system with
	states drawn from a Gaussian distribution, with small standard deviation ($\sim 0.01$),
	centered around zero. (Bottom panel) time evolution of the elements in the array of size
	$N=100$, where only one element has $a_1=1$ (denoted by arrow) and $99$ elements have
	$a_i=0$.} \label{biasN}

\end{figure}

One may wonder if the fundamental one-bit detection limit can be achieved in our system. In
Fig.~\ref{biasN}, we have shown how a large system, $N=1000$, responds to only {\em one}
distinct element in the array. It demonstrates that there exist a range of global bias which
allows the system to yield $P = 0$ for $N_1 = 0$ and $P = 1$ when $N_1 = 1$. So by tuning the
global bias we can obtain a system where a single $a_i=1$ can draw the whole system to the
upper well.

Specifically then, in the representative example displayed in Fig.\ref{biasN}, in a lattice of
$100$ elements, when all $a_i=0$ the mean field is $\sim -1$, reflecting the fact that all
elements go to the lower well, as expected. However, when {\em one} of the $a_i$ is $1$ (i.e.
$N_1 = 1, N_0 = 99$), the mean field evolves to $\sim 1$, reflecting the fact that all elements
have been attracted to the upper well now, driven by this one different element. So even though
only one element in the array would have evolved to the upper well in the uncoupled case, when
strongly coupled, all $100$ elements are dragged rapidly to the upper well.

\begin{figure}[t]
\begin{center}
\includegraphics[width=0.5\linewidth,angle=270]{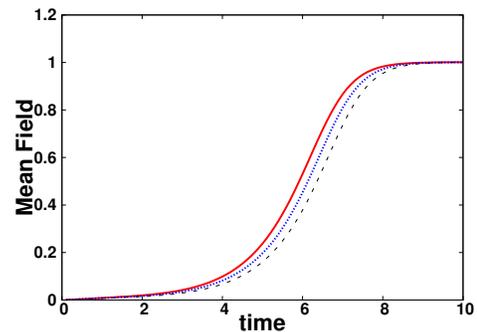}
\end{center}

\caption{ Evolution of the mean field as a function of time for different array sizes: $100$
(red solid), $1000$ (black dashed) and $10000$ (blue dotted), with $N_1/N = 0.01$ and bias $b =
-0.005$.  Clearly, the time needed to reach the attracting state does not depend on $N$.}
\label{size}

\end{figure}

One can analyze the dynamics by considering $\langle x \rangle = x_{av} + \eta$, where $\eta$
is a random fluctuation about the true thermodynamic average $x_{av}$ ($\sim 0$), with $\eta
\sim 1/\sqrt{N}$. So the effective dynamics of each element is:
\begin{equation}
\dot{x_i} = (1-C) x_i - x_i^3 + a_i + b + C x_{av} + \frac{C}{\sqrt{N}} \eta = F(x_i) + D \eta
\end{equation}
The probability $P(x)$ of obtaining the system in state $x$ for the elements, can be analyzed
by solving for the steady state distribution arising from the relevant Fokker Planck equation,
namely $P(x)= A \exp(-2 \phi(x)/D)$ where $A$ is a normalization constant, $D= C/\sqrt{N}$ is
the strength of the random fluctuations, and $-\partial \phi(x)/\partial x = F(x)$
\cite{hasty}.  Using this we verify that the uncoupled symmetric case, i.e. with $C=0$ and $b =
0$, yields the following: elements with $a_i=0$ yield equal probability of residence in either
of the two wells (centered around $1$ and $-1$), and for elements with $a_i=1$ the probability
$P(x)$ shifts entirely to the upper well ($\sim 1$). When there is no coupling or very weak
coupling, this is indeed the case in our simulations as well.

However, for strong coupling ($C \sim 1$), we have a very different scenario: when all $a_i=0$
(with $N \rightarrow \infty$) and even the slightest negative bias, $P(x)$ peaks sharply in a
well with $x < 0$. That is, the entire system is synchronized and attracted to the lower well.
Contrast this to the unsynchronized situation in weak coupling, where the elements go to either
upper or lower well depending on their initial state. Similarly, for strong coupling, even
slight positive bias drives all the elements to the upper well.

Now consider the strong coupling case with a few $a_i = 1$. The elements with $a_i =1$ have
$P(x)$ centered sharply at a well $\sim 1$ (as $a_i$ acts as a positive bias). As these $N_1$
elements evolve to the upper well, $x_{av}$ becomes slightly positive, and even elements with
$a_i=0$ experience a positive bias: $b + C x_{av} > 0$, which shifts $P(x)$ entirely to a well
at $x > 0$. Following the small initial positive push, there is a strong positive feedback
effect that drives the $x_{av}$ to more and more positive values, and consequently the stable
attracting well shifts rapidly towards $\sim x^*_+$.

One can also rationalize this mechanism intuitively as follows: when the initial system has $x
\sim 0$, namely the system is poised on the ``barrier'' between the two wells, the state is
tipped to the well at $x^*_{+}$ if $\dot{x} >0$ and to the lower well at $x^*_{-}$ if $\dot{x}
< 0$. Now, initial $F(x) \sim 0$ as the system is at the unstable maximum of the potential, and
$\dot{x_i} \sim a_i + b + (\langle x \rangle -x_i)$, where $b$ is close to $0$, and $\langle x
\rangle -x_i$ is small in magnitude. So for elements with $a_i=1$, $\dot{x} \sim 1$, and for
$a_i = 0$, $\dot{x}$ is also positive, though small in magnitude, as $x_i < \langle x \rangle$.
After this infinitesimal initial push towards $x^*_{+}$, all elements evolve rapidly towards
that stable upper well, as $F(x)$ gets increasingly positive.

{\em Robustness of the phenomena:} In order to gauge the generality of our observations, we
have considered different nonlinear functions $F(x)$ in Eq.(1). For example, we explored a
system of considerable biological interest, namely, a system of coupled synthetic gene
networks.  We used the quantitative model, developed in \cite{hasty}, describing the regulation
of the operator region of $\lambda$ phase, whose promoter region consists of three operator
sites. The chemical reactions describing this network, given by suitable rescaling yields
\cite{hasty} $$F_{gene}(x) = \frac{m (1 + x^2 + \alpha \sigma_1 x^4)}{1 + x^2 + \sigma_1 x^4 +
\sigma_1 \sigma_2 x^6} - \gamma_x x$$ where $x$ is the concentration of the repressor. The
nonlinearity in this $F(x)$ leads to a double well potential, and different $\gamma$ introduces
varying degrees of asymmetry in the potential. We studied a system of coupled genetic
oscillators given by: $\dot{x_i} = F_{gene}(x_i) + C (\langle x \rangle -x_i) + a_i + b$, where
$C$ is the coupling strength, and $b$ is a small global bias.  We observe similar features in
this system as well.

In addition, we studied various different coupling forms. For instance, a system of $N$ coupled
nonlinear systems, where the evolution of element $i$ is given by:
\begin{equation}
\dot{x_i} = F(x_i) + a_i + C \langle x \rangle + b
\label{array2}
\end{equation}
where $C$ is the coupling strength and $\langle x \rangle$ is the mean field given by
Eq.(\ref{xav}). Furthermore, we considered small world networks, where varying sets of regular
links were replaced by random connections. Lastly, we explored networks with different ranges
of coupling, namely the coupling occured over increasingly large subsets of neighbours, up to
the global coupling limit. Qualitatively, the same ultra-sensitivity to heterogeneity has been
observed for all these different dynamical systems and coupling forms.

{\em Relevance to Information Search:} Lastly, we address a problem of database searching,
utilizing these strongly coupled bistable dynamical systems as the building block of the search
engine. We propose a method, involving a single global operation, to determine the existence of
very few specified items in a given, arbitrarily large, unsorted database.

First we use the bistable elements to stably encode $N$ binary items ($0$ or $1$) by setting
$a_i$ ($i = 1, \dots N$) to take values $0$ or $1$, respectively [see Fig.1(a)]. This creates a
(unsorted) binary database. Then, using the scalable ultra-sensitivity demonstrated above, one
can search this arbitrarily large database for the existence of a single different bit (say a
single $1$ in a string of $0$'s) by making just one measurement of the evolved mean field of
the whole array.

Furthermore, we have a look-up table relating critical $N_1$ to global bias $b$ (cf. Fig.
\ref{biasN}).  So by sweeping $b$ we can find where the cross-over to the upper well occurs.
This average critical value can be used to gauge the {\em number of ones} present in the system
as well.

The significant feature here that allows this massive parallelization, is the fact that the
time taken to reach the attracting mean field value {\em does not scale with system size} (see
Fig.~\ref{size}). In fact the time taken to reach the mean field that encodes the output is
{\em independent} of $N$.

Another important feature of this scheme is that it employs a {\em single} simple global
operation, and does not entail accessing each item separately at any stage. In comparison, for
example, a conventional search algorithm with binary encoding will take $O(log N)$ procedural
steps for binary search of an ordered tree \cite{search}. In addition, there is the time
required for ordering, which typically scales as $O (N \times log N)$.  Alternate ideas to
implement the increasingly important problem of search have included the use of quantum
computers \cite{grover}, which involves scaling of time steps as $O(\sqrt{N})$.

In summary, the collective response of $N$ globally coupled bistable elements can strongly
reflect the presence of very few non-identical elements in a very large array of otherwise
identical elements. Counter-intuitively, the mean field evolves to the stable state of the
minority population, rather than that of the bulk of the array.  Adjusting the global bias
enables us to observe robust {\em one bit} sensitivity to diversity in this array.  Further,
the time needed to reach the attracting state does not increase with system size. Thus this
phenomenon has much relevance to the problem of massively parallelized search. Lastly, this
scalable ultra-sensitivity is a generic and robust phenomenon, and can potentially be observed
in social and biological networks \cite{scheffer}, coupled nano-mechanical resonators
\cite{nano}, and coupled laser arrays \cite{laser}.

\end{document}